\documentclass[journal]{IEEEtran}
\usepackage[colorlinks,linkcolor=blue,anchorcolor=blue,citecolor=blue]{hyperref}
\usepackage{cite}
\usepackage{url}
\usepackage[caption=false]{subfig}
\usepackage{float}
\usepackage{makecell}

\usepackage{amsmath,amssymb,amsfonts}
\usepackage{hyperref}
\allowdisplaybreaks[4]
\usepackage{graphicx}
\usepackage{enumerate}
\usepackage{array}
\usepackage{multirow}
\usepackage{color}
\usepackage{colortbl}
\usepackage[ruled,vlined,linesnumbered]{algorithm2e}
\usepackage{tikz}

\begin{document}

\title{Stochastic Model Predictive Control under AC Power-Flow Constraints Using Generative Learning}

\author{Jie Zhu,~\IEEEmembership{Graduate Student Member,~IEEE,} Yinliang Xu,~\IEEEmembership{Senior Member,~IEEE}, and Guan Wang 
\thanks{
This work has been submitted to the IEEE for possible publication. Copyright may be transferred without notice, after which this version may no longer be accessible.

Jie Zhu and Yinliang Xu are with the Tsinghua Shenzhen International Graduate School, Tsinghua University, Shenzhen 518055, China (e-mail: zhuj24@mails.tsinghua.edu.cn; xu.yinliang@sz.tsinghua.edu.cn, corresponding author).

Guan Wang is with the Department of Applied Mathematics, Hong Kong Polytechnic University, Hong Kong, China. (e-mail: guan1.wang@polyu.edu.hk)

}}
\markboth{IEEE TRANSACTIONS ON POWER SYSTEMS}%
{Shell \MakeLowercase{\textit{et al.}}: A Sample Article Using IEEEtran.cls for IEEE Journals}

\IEEEpubid{}

\maketitle

\begin{abstract}
This paper proposes an end-to-end generative framework for efficiently solving multi-period and multi-scenario stochastic model predictive control (SMPC) problems under nonlinear AC power-flow constraints. Conventional deterministic neural surrogates rely on a single-shot prediction, which makes reliable feasibility difficult to achieve because one dispatch trajectory must simultaneously satisfy nonlinear constraints across all scenarios and time periods. To address this limitation, a conditional stochastic neural generator (CSNG) is developed to produce multiple candidate dispatch trajectories for each uncertainty instance, enabling the recovery of a feasible and economical solution through candidate selection. A feasibility-aware self-supervised distribution-shaping scheme is further introduced to promote constraint satisfaction, candidate diversity, and operating economy without requiring computationally expensive SMPC solution labels, while mitigating candidate collapse in generative ACOPF learning. To support efficient end-to-end training, a constraint-aware differentiable architecture is introduced. It employs a projection mechanism to exactly enforce box and ramping constraints while preserving informative gradients near active bounds, together with a differentiable equality-completion surrogate for efficient AC power-flow reconstruction. Case studies on the IEEE 14- and 118-bus systems demonstrate $100\%$ feasibility, optimality gaps below $2\%$, and computational efficiency suitable for intraday dispatch. The implementation is publicly available at \url{https://github.com/JieZhu6/Generative_SMPC}.
\end{abstract}


\section{Introduction}
\IEEEPARstart{T}{he} increasing penetration of renewable energy introduces substantial uncertainty into nodal net injections, posing challenges for intraday power-system operation. Model predictive dispatch (MPC) is well suited to this setting, as it repeatedly updates dispatch decisions over a receding horizon using the latest forecasts while accounting for network and intertemporal constraints. To explicitly account for forecast uncertainty, various uncertainty-aware MPC formulations have been developed based on chance-constrained \cite{10762839}, robust \cite{11595584}, and distributionally robust optimization \cite{10089877}. However, tractable implementations often rely on DC power-flow (PF) models or convex approximations, as uncertainty-aware reformulations are difficult under nonlinear and nonconvex AC PF constraints. Scenario-based stochastic MPC (SMPC) provides a flexible alternative by directly enforcing constraints over sampled uncertainty realizations while retaining the full AC model \cite{10271696}. However, the resulting SMPC problem remains computationally demanding, with complexity growing rapidly with the prediction horizon and number of scenarios, which limits real-time intraday application.

Learning-based optimization has emerged as a promising approach to reduce the computational burden of repeatedly solving power-system optimization problems. By using neural networks (NNs) as surrogate optimizers, it approximates the mapping from problem parameters to high-quality dispatch decisions, shifting most computation offline and enabling fast online inference. A classical paradigm relies on supervised learning to imitate optimal solutions obtained from conventional solvers \cite{pan2022deepopf,nellikkath2022physics}. Alternatively, fully self-supervised approaches avoid the need for optimal labels by constructing the training loss directly from the economic objective and penalties for physical and operational constraint violations \cite{huang2024unsupervised,10540179}. However, extending NN-based surrogate optimization to SMPC introduces an additional scalability challenge. If all uncertainty scenarios are used directly as inputs, the input dimension grows rapidly with the number of scenarios, limiting the scalability of the neural architecture. To address this issue, Chen et al.~\cite{10965352} map observable auxiliary information used for uncertainty forecasting directly to SMPC dispatch decisions, thereby avoiding an explicit representation of the scenario set. Zhou et al.~\cite{11192606} further exploit the partial permutation invariance of uncertainty scenarios and develop a scenario-embedding architecture \cite{patel2022neur2sp} to obtain a compact representation, mitigating the dimensional growth associated with large scenario sets.

Beyond neural architectures and training paradigms, solution feasibility is critical to reliable NN-based optimization. For box-constrained variables, existing methods commonly use smooth bounded mappings, such as sigmoid or tanh, or hard clipping to map raw NN outputs into admissible ranges \cite{9844847}. However, smooth mappings suffer from vanishing gradients near active bounds, while hard clipping blocks gradients outside the feasible range \cite{11660390}. Both effects can weaken the optimization signal near or beyond active constraints, thereby impairing surrogate training. For the remaining constraints, existing studies \cite{pan2022deepopf,nellikkath2022physics,huang2024unsupervised,10540179,10965352,11192606,patel2022neur2sp} mainly rely on soft penalties, which promote but do not guarantee feasibility at inference. To improve feasibility, several studies introduce explicit repair or projection mechanisms. Han et al.~\cite{10411984} recover feasibility by solving an auxiliary projection problem with a numerical solver at inference, but the additional optimization overhead can substantially reduce computational efficiency, particularly for large-scale SMPC problems. Donti et al.~\cite{donti2021dc3} avoid external solvers by unrolling iterative feasibility-correction steps into the learning pipeline; however, reliable constraint satisfaction remains difficult for nonconvex problems. Chen et al.~\cite{10256159} introduce a repair layer to restore power-balance consistency, but the resulting correction does not fully account for network-constrained AC feasibility. Liang et al.~\cite{liang2024homeomorphic} propose a homeomorphic transformation with bisection-based projection, whose applicability relies on simple connectivity of the feasible region. To relax this restriction, Liang et al.~\cite{liangefficient} further employ an NN-predicted interior point to construct projection paths for more general feasible sets; nevertheless, the effectiveness of the projection depends strongly on the accuracy of the predicted interior point, which becomes difficult to maintain in high-dimensional nonconvex problems.

Taken together, these limitations make reliable feasibility recovery particularly challenging for SMPC. A fundamental issue is that the aforementioned methods still rely on a single deterministic prediction. Since each dispatch trajectory must satisfy nonlinear constraints across multiple scenarios and time periods. Once this occurs, the single-shot surrogate provides no alternative candidate for the same problem instance. Generative learning offers a different paradigm by modeling an input-conditioned solution distribution rather than a single-valued input--solution mapping \cite{11482518}. Recent studies have explored this idea for nonconvex optimization using different generative architectures. Liang and Chen~\cite{liang2024generative} employ Rectified Flow to learn conditional distributions of high-quality solutions. DiffOPF~\cite{hoseinpour2027diffopf} applies diffusion models to AC-OPF and learns load-to-dispatch distributions from labeled operating data generated under perturbed generator costs. Ding et al.~\cite{ding2025diffusion} further introduce weighted bootstrapped refinement to improve the alignment between generated solutions and the feasible region. Meng et al.~\cite{mengsolving} combine flow matching with reinforcement learning to progressively refine constrained solutions. These studies demonstrate the potential of generative models for nonconvex optimization.

In contrast to existing works that use generative models to represent multi-valued solution mappings, this work exploits generative models to improve feasible-solution recovery in SMPC. For each uncertainty instance, multiple dispatch trajectories provide multiple opportunities to identify a feasible and economical solution. However, extending existing generative solvers to SMPC faces two challenges. First, existing methods \cite{liang2024generative,hoseinpour2027diffopf,ding2025diffusion,mengsolving} rely on high-quality labeled solutions, which are costly to obtain for multi-period and multi-scenario SMPC. Second, solution multiplicity can be limited in standard AC-OPF instances \cite{bukhsh2013local}, causing generated candidates to concentrate in a narrow region without explicit diversity preservation and thus reducing the benefit of generative models.

To address these challenges, this work exploits generative model as a feasible-solution recovery mechanism for SMPC. A conditional stochastic neural generator (CSNG) maps the SMPC problem parameters to multiple candidate dispatch trajectories and identifies a feasible and economical solution through candidate selection. The main contributions are summarized as follows:

1) A generative learning paradigm is developed for SMPC by exploiting stochastic generation to identify feasible and economical solutions. Unlike existing single-shot approaches \cite{10965352,11192606}, the proposed CSNG generates multiple dispatch candidates for each uncertainty instance and identifies a feasible solution through candidate selection, rather than relying on a single deterministic prediction.

2) A feasibility-aware distribution-shaping self-supervised training framework is proposed to directly optimize the SMPC objective and constraints. Unlike existing generative methods that rely on high-quality solution labels \cite{liang2024generative,hoseinpour2027diffopf,ding2025diffusion,mengsolving}, it avoids costly SMPC labeling while guiding candidates toward diverse, feasible, and economical regions.

3) A constraint-aware differentiable architecture is developed for efficient feasibility-aware learning. It exactly enforces box and ramping constraints while mitigating the vanishing gradients of smooth bounded mappings near active bounds and the gradient blocking of hard clipping outside the feasible range. Moreover, it avoids the repeated PF solves and Jacobian-based linear systems required by implicit-differentiation approaches \cite{donti2021dc3,10256159,liang2024homeomorphic,liangefficient} through minimal-state learning and analytical PF reconstruction.

The remainder of this paper is organized as follows. Section~\ref{sec:2} formulates the SMPC problem. Section~\ref{sec:method} presents the proposed CSNG formulation and training framework. Section~\ref{sec:case} reports numerical case studies to evaluate the effectiveness of the proposed method. Finally, Section~\ref{sec:con} concludes the paper.

\section{Problem Formulation}
\label{sec:2}
Consider a transmission network with bus set $\mathcal{N}$, conventional-generator bus set $\mathcal{N}_g$, and reference bus $r_0$. Let $\mathcal N_{\mathrm{PV}}$ denotes the sets of P-V buses, with $r_0$ excluded from $\mathcal N_{\mathrm{PV}}$. An SMPC scheme is considered over the prediction horizon $\mathcal{T}=\{1,\ldots,T\}$ with scenario set $\mathcal{N}_s$. The resulting SMPC problem is formulated as
\begin{subequations}
\label{eq:smpc_acopf}
\begin{align}
&\min~\sum_{t\in\mathcal{T}}\sum_{i\in\mathcal{N}_\mathrm{PV}}
\left[a_i \left(P^g_{i,t}\right)^2+ b_i P^g_{i,t}+ c_i\right]
\nonumber\\
& + \frac{1}{|\mathcal{N}_s|}\sum_{s\in\mathcal{N}_s}\sum_{t\in\mathcal{T}}
  \left[a_i \left(P^g_{s,r_0,t}\right)^2+ b_i P^g_{s,r_0,t}+ c_{r_0}\right]
\label{eq:smpc_acopf_obj}\\
&\text{s.t.}\;
\begin{aligned}[t]
P^g_{s,i,t}&-P^d_{s,i,t}
=V_{s,i,t}\sum_{j\in\mathcal{N}}V_{s,j,t}\\
&\times\bigl(
G_{ij}\cos\theta_{s,ij,t}
+B_{ij}\sin\theta_{s,ij,t}
\bigr),
~\forall s\in\mathcal{N}_s
\end{aligned}
\label{eq:smpc_acopf_pbal}\\
&
\begin{aligned}[t]
Q^g_{s,i,t}&-Q^d_{s,i,t}
=V_{s,i,t}\sum_{j\in\mathcal{N}}V_{s,j,t}\\
&\times\bigl(
G_{ij}\sin\theta_{s,ij,t}
-B_{ij}\cos\theta_{s,ij,t}
\bigr),
~\forall s\in\mathcal{N}_s
\end{aligned}
\label{eq:smpc_acopf_qbal}\\
&
\begin{aligned}[t]
P_{s,ij,t}
&=-G_{ij}V_{s,i,t}^2+V_{s,i,t}V_{s,j,t}\\
&\times\bigl(
G_{ij}\cos\theta_{s,ij,t}
+B_{ij}\sin\theta_{s,ij,t}
\bigr),
~\forall s\in\mathcal{N}_s
\end{aligned}
\label{eq:smpc_acopf_pflow}\\
&
\begin{aligned}[t]
Q_{s,ij,t}
&=B_{ij}V_{s,i,t}^2+V_{s,i,t}V_{s,j,t}\\
&\times\bigl(
G_{ij}\sin\theta_{s,ij,t}
-B_{ij}\cos\theta_{s,ij,t}
\bigr),
~\forall s\in\mathcal{N}_s
\end{aligned}
\label{eq:smpc_acopf_qflow}\\
&
|S_{s,ij,t}|^2
=P_{s,ij,t}^2+Q_{s,ij,t}^2,
~\forall s\in\mathcal{N}_s
\label{eq:smpc_acopf_sflow}\\
&
|S_{s,ij,t}|
\leq\overline{S}_{ij},
~\forall s\in\mathcal{N}_s
\label{eq:smpc_acopf_slimit}\\
&
\underline{V}_i
\leq V_{s,i,t}
\leq\overline{V}_i,\quad
\underline{\theta}_{ij}
\leq\theta_{s,ij,t}
\leq\overline{\theta}_{ij},
~\forall s\in\mathcal{N}_s
\label{eq:smpc_acopf_vanglelimit}\\
&
\underline{P}^g_i
\leq P^g_{s,i,t}
\leq\overline{P}^g_i,\quad
\underline{Q}^g_i
\leq Q^g_{s,i,t}
\leq\overline{Q}^g_i,
~\forall s\in\mathcal{N}_s
\label{eq:smpc_acopf_genlimit}\\
&
-R_i^{\mathrm{down}} \leq P^g_{s,i,t}-P^g_{s,i,t-1} \leq R_i^{\mathrm{up}},
~\forall s\in\mathcal{N}_s
\label{eq:smpc_acopf_ramp}\\
&
\theta_{s,r_0,t}=0,
~\forall s\in\mathcal{N}_s
\label{eq:smpc_acopf_ref}
\end{align}
\end{subequations}
where $P_{s,i,t}^{d}$ and $Q_{s,i,t}^{d}$ denote the net active and reactive demands, i.e., load demand minus renewable generation, at bus $i$ and time $t$ under scenario $s$. The variables $V_{s,i,t}$ and $\theta_{s,i,t}$ denote the corresponding voltage magnitude and phase angle, with $\theta_{s,ij,t}=\theta_{s,i,t}-\theta_{s,j,t}$. For each P-V bus $i\in\mathcal N_{\mathrm{PV}}$, the active-power dispatch $P_{i,t}^{g}$ and voltage-magnitude set point $V_{i,t}$ are shared across all scenarios and therefore carry no scenario index. The variables $P_{s,ij,t}$, $Q_{s,ij,t}$, and $S_{s,ij,t}$ denote the active, reactive, and apparent power flows from bus $i$ to bus $j$, respectively. The parameters $G_{ij}$ and $B_{ij}$ are the real and imaginary parts of the $(i,j)$th entry of the nodal admittance matrix. The coefficients $a_i$, $b_i$, and $c_i$ specify the generation cost. The parameters $\underline{P}^g_i$ and $\overline{P}^g_i$ denote the active-power limits, while $\underline{Q}^g_i$ and $\overline{Q}^g_i$ denote the reactive-power limits. The parameters $\underline{V}_i$ and $\overline{V}_i$ are the voltage-magnitude limits, and $\underline{\theta}_{ij}$ and
$\overline{\theta}_{ij}$ are the phase-angle-difference limits. The parameter $\overline{S}_{ij}$ is the apparent-power thermal limit. Furthermore, $R_i^{\mathrm{up}}$ and $R_i^{\mathrm{down}}$ are the ramp-up and ramp-down limits, respectively.

\section{Methodology}
\label{sec:method}
\begin{figure*}[!t]
   \centering
   \includegraphics[width=0.95\textwidth]{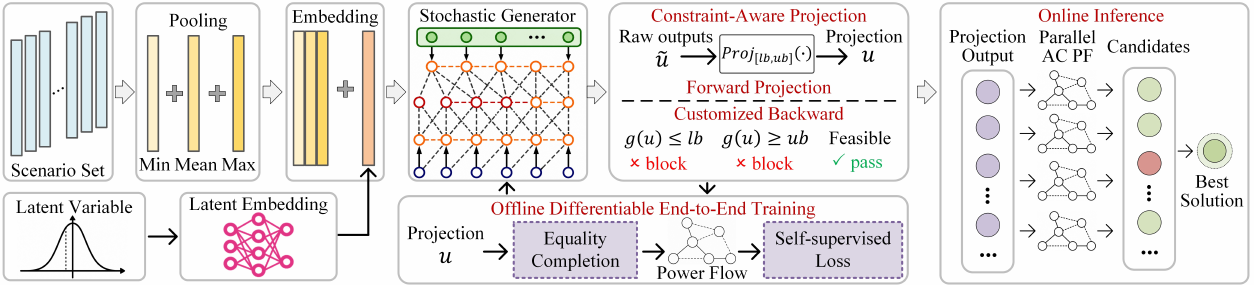}
   \caption{Schematic Architecture of the CSNG}
   \label{fig:framework}
\end{figure*}

Directly solving the nonconvex SMPC problem in \eqref{eq:smpc_acopf} becomes computationally demanding as the prediction horizon and number of uncertainty scenarios increase. The proposed framework in Fig.~\ref{fig:framework} addresses this challenge through three components. First, a CSNG maps a compact scenario representation and latent variables to multiple candidate dispatch trajectories. Second, a feasibility-aware self-supervised scheme guides the candidates toward diverse, feasible, and economical regions without optimal labels. Third, a constraint-aware differentiable architecture enforces box and ramping constraints and recovers the remaining AC network states through a differentiable surrogate. Together, these components enable efficient end-to-end learning for SMPC.

\subsection{Conditional Stochastic Neural Generator}
Let $C_{obj}(\cdot)$ denote the objective function in \eqref{eq:smpc_acopf}. The SMPC problem can then be compactly formulated as follows:
\begin{subequations}
\label{eq:compact_formulation}
\begin{align}
\min_{\mathbf{y}}\quad
&C_{obj}\left(\mathbf{y}\right)
\label{eq:scenario_value_obj}
\\
\mathrm{s.t.}\quad
&\mathbf{h_{eq}}\left(\mathbf{y};\boldsymbol{\xi}_s\right)=\mathbf{0},~\forall s\in\mathcal{N}_s
\label{eq:scenario_value_eq}
\\
&\mathbf{g_{ineq}}\left(\mathbf{y};\boldsymbol{\xi}_s\right)\leq\mathbf{0},~\forall s\in\mathcal{N}_s
\label{eq:scenario_value_ineq}
\end{align}
\end{subequations}
where $\mathbf{y}$ denotes the decision variables, and $\boldsymbol{\xi}_s$ denotes the load uncertainty trajectory associated with scenario $s$. The function $\mathbf{h_{eq}}(\cdot)$ collects the power balance equations and branch-flow equations, while $\mathbf{g_{ineq}}(\cdot)$ collects the corresponding generator-output limits, voltage limits, phase-angle limits, transmission-line thermal limits, and ramping constraints. 

\label{subsec:stochastic_generator}
\subsubsection{Motivation for Generative Modeling}
\label{subsubsec:generative_motivation}
Most existing learning-based optimization methods employ a deterministic NN as a surrogate for solving \eqref{eq:compact_formulation}. However, such deterministic neural solvers typically provide only one candidate solution for each problem instance. Once the single prediction is infeasible, no alternative candidate is available for the same instance. This observation motivates the use of a conditional stochastic neural generator (CSNG). Instead of producing a single deterministic solution, the CSNG $G_{\sigma}(\cdot)$ parameterized by $\sigma$ introduces a latent random variable $\mathbf z$ and conceptually represents
\begin{equation} \mathbf y=G_{\sigma}\left(\boldsymbol{\xi},\mathbf z\right),\quad \mathbf z\sim p_{\mathbf z}
\label{eq:conditional_generation} \end{equation}
where $p_{\mathbf z}=\mathcal N(\mathbf 0,\mathbf I)$ is the latent distribution. For a fixed SMPC instance, different realizations of $\mathbf z$ produce different candidate dispatch trajectories.

Let $\mathcal U_{\boldsymbol{\xi}}^{\mathrm{fea}}$ denote the set of dispatch trajectories whose corresponding AC PF realizations satisfy all physical and operational constraints. The feasibility probability of a single generated candidate is defined as
\begin{equation} p_{\mathrm{fea}}(\boldsymbol{\xi};\theta)=\Pr_{\mathbf z\sim p_{\mathbf z}}\left[G_{\sigma}(\boldsymbol{\xi},\mathbf z)\in\mathcal U_{\boldsymbol{\xi}}^{\mathrm{fea}}\right].
\label{eq:single_feasibility_probability} \end{equation}

For a given SMPC instance, $K$ independent latent samples are drawn as
\begin{equation} \mathbf z^{(k)}\overset{\mathrm{i.i.d.}}{\sim}p_{\mathbf z},\quad k=1,\ldots,K,
\label{eq:latent_samples} \end{equation}
and the corresponding candidate dispatch trajectories are
\begin{equation} \mathbf y^{(k)}=G_{\sigma}\left(\boldsymbol{\xi},\mathbf z^{(k)}\right),\quad k=1,\ldots,K.
\label{eq:parallel_candidates} \end{equation}

Under independent latent sampling, the probability that at least one candidate belongs to $\mathcal U_{\boldsymbol{\xi}}^{\mathrm{fea}}$ is
\begin{equation} P_{\mathrm{hit}}^{(K)}(\boldsymbol{\xi};\theta)=1-\left[1-p_{\mathrm{fea}}(\boldsymbol{\xi};\theta)\right]^K.
\label{eq:feasibility_amplification} \end{equation}
Equation~\eqref{eq:feasibility_amplification}  highlights the main advantage of generative modeling in the present setting: multiple candidates provide multiple opportunities to recover a feasible solution for the same SMPC instance. At inference, the generated candidates are first screened according to feasibility, and the lowest-cost candidate is subsequently selected from the feasible subset.

\subsubsection{Generator Architecture}
\label{subsubsec:generator_architecture}

The representation in \eqref{eq:parallel_candidates} is conceptual, as directly concatenating all scenario realizations would make the generator input dimension grow linearly with $|\mathcal N_s|$. To obtain a compact fixed-dimensional representation, empirical pooling is applied over the scenario dimension using element-wise minimum, mean, and maximum operations:
\begin{equation}
\mathbf \xi_{\mathrm{pool}}=
\left(\displaystyle\min_{s\in\mathcal N_s}\boldsymbol{\xi}_s,\displaystyle
\frac{1}{|\mathcal N_s|}\sum_{s\in\mathcal N_s}\boldsymbol{\xi}_s
,\displaystyle\max_{s\in\mathcal N_s}\boldsymbol{\xi}_s\right)\label{eq:scenario_pooling_concat}
\end{equation}
where the minimum and maximum are evaluated component-wise. The pooled representation captures the range and central tendency of the uncertainty scenarios while remaining independent of $|\mathcal N_s|$. To capture intertemporal dependencies in multi-period dispatch, the stochastic generator is implemented using a non-causal temporal convolutional network (TCN), allowing each decision to exploit uncertainty information over the entire prediction horizon while preserving temporal structure. Accordingly, the pooled representation in \eqref{eq:scenario_pooling_concat} is arranged as the temporal sequence
\begin{equation} \mathbf \xi_{\mathrm{pool}}=\begin{bmatrix}(\mathbf \xi_{1}^{\mathrm{pool}})^{\top}&\cdots&(\mathbf \xi_{T}^{\mathrm{pool}})^{\top}\end{bmatrix}^{\top}\in\mathbb R^{T\times 3d_{\xi}}\label{eq:temporal_pooled_sequence} \end{equation}
where $\mathbf \xi_t^{\mathrm{pool}}\in\mathbb R^{3d_{\xi}}$ denotes the minimum--mean--maximum pooled uncertainty feature at time $t$, and $d_{\xi}$ denotes the dimension of the uncertain quantities at each time period.

For each candidate trajectory, a latent vector $\mathbf z\sim\mathcal N(\mathbf 0,\mathbf I)$ is sampled and shared across all time periods. This trajectory-level latent variable enables different realizations of $\mathbf z$ to represent distinct dispatch trajectories. The latent vector is first embedded and then concatenated with the pooled uncertainty feature at each time step:
\begin{equation} \mathbf h_t^{(0)}=\phi_{\mathrm{in}}\left(\mathbf \xi_t^{\mathrm{pool}},\phi_z(\mathbf z)\right),\quad t\in\mathcal T\label{eq:tcn_input_embedding} \end{equation}
where $\phi_z(\cdot)$ denotes the learnable latent embedding, $\phi_{\mathrm{in}}(\cdot)$ denotes the input embedding, and $\mathbf h_t^{(0)}$ is the resulting input feature of the TCN at time $t$.

The embedded sequence is then processed by stacked non-causal dilated convolutional blocks. At the $\ell$th layer, the convolutional feature at time $t$ is computed as
\begin{equation} \mathbf c_t^{(\ell)}=\sum_{q=-m}^{m}\mathbf W_q^{(\ell)}\mathbf h_{t+qd_{\ell}}^{(\ell-1)}+\mathbf b^{(\ell)},\quad k_c=2m+1\label{eq:noncausal_dilated_conv} \end{equation}
where $k_c$ is the convolution kernel size, $m=(k_c-1)/2$ specifies the temporal neighborhood on each side of $t$, $d_{\ell}$ is the dilation factor of layer $\ell$, and $\mathbf W_q^{(\ell)}$ and $\mathbf b^{(\ell)}$ denote the convolution weights and bias, respectively. $\mathbf h_t^{(\ell-1)}$ denotes the input feature to the $\ell$th block, while $\mathbf c_t^{(\ell)}$ is subsequently passed through the nonlinear and residual operations of the block to obtain $\mathbf h_t^{(\ell)}$. Symmetric padding is employed to preserve the sequence length.

Because \eqref{eq:noncausal_dilated_conv} incorporates information from both sides of time $t$, stacking dilated convolutional blocks progressively enlarges the temporal receptive field. The dilation configuration is selected such that the final receptive field covers the entire prediction horizon. Denoting the resulting non-causal TCN by $\operatorname{NCTCN}_{\sigma_h}(\cdot)$, where $\sigma_h$ collects its trainable parameters, the horizon-wide representation is written compactly as
\begin{equation} \mathbf H=\operatorname{NCTCN}_{\sigma_h}\left(\mathbf \xi_{\mathrm{pool}},\mathbf z\right) \label{eq:nctcn_representation} \end{equation}
where $\mathbf H$ denotes the resulting temporal feature sequence that jointly encodes the pooled uncertainty information and the latent realization.

On the output side, directly generating the complete operating point $\mathbf y$ would require the CSNG to produce samples satisfying the nonlinear PF equality constraints. To avoid imposing this burden on the generator, the variables are partitioned as
\begin{equation}
\mathbf y
=
\begin{bmatrix}
\mathbf u\\
\mathbf w
\end{bmatrix}
\label{eq:variable_partition}
\end{equation}
where $\mathbf u$ contains the free decision variables and $\mathbf w$ contains the dependent variables determined by the AC PF equations. Specifically, the free-variable vector consists of the active-power set points at PV buses and the voltage-magnitude set points at PV buses and the reference bus, i.e., $\mathbf u=\left(\mathbf P_{\mathrm{PV}}^{g},\mathbf V_{\mathrm{PV}},\mathbf V_{r_0}\right)$. The remaining system states are recovered from the AC PF equations for the given uncertainty realization. Accordingly, the free variables are generated as
\begin{equation}
\mathbf u=G_{\sigma}^{u}\left(\mathbf \xi_{\mathrm{pool}},\mathbf z\right)=\phi_{\mathrm{out},\sigma_o}
\left(\operatorname{NCTCN}_{\sigma_h}\left(\mathbf \xi_{\mathrm{pool}},\mathbf z\right)\right).
\label{eq:nctcn_generator}
\end{equation}
where $\phi_{\mathrm{out},\sigma_o}(\cdot)$ denotes the output mapping that transforms the temporal feature sequence $\mathbf H$ into the free-variable trajectory, and $\sigma=\{\sigma_h,\sigma_o\}$ collects the trainable parameters of the NCTCN backbone and the output mapping.

Given $\mathbf u$ and the full scenario realization $\boldsymbol{\xi}$, the corresponding operating point is completed through
\begin{equation}
\mathbf y=\begin{bmatrix}\mathbf u\\\mathbf w\end{bmatrix}
=\begin{bmatrix}G_{\sigma}^{u}\left(\mathbf \xi_{\mathrm{pool}},\mathbf z\right)\\[1mm]\boldsymbol{\psi}_{\mathrm{pf}}\left(G_{\sigma}^{u}
\left(\mathbf \xi_{\mathrm{pool}},\mathbf z\right);
\boldsymbol{\xi}\right)
\end{bmatrix}
\label{eq:y_completion}
\end{equation}
where $\boldsymbol{\psi}_{\mathrm{pf}}(\cdot)$ denotes the equality-completion mapping induced by the AC PF equations.

\subsubsection{Constraint-Aware Projection}
\label{subsubsec:one_sided_projected_backward}
The structure of $\mathbf u$ also allows several operational constraints to be enforced directly at the generator output. Let
\begin{equation}
\widetilde{\mathbf u}=\left(\widetilde{\mathbf P}_{\mathrm{PV}}^{g},\widetilde{\mathbf V}_{\mathrm{PV}},\widetilde{\mathbf V}_{r_0}\right)
=
\widetilde{G}_{\sigma}^{u}
\left(\boldsymbol{\xi}_{\mathrm{pool}},\mathbf z\right)
\label{eq:raw_generator}
\end{equation}
denote the raw output of the stochastic neural generator. Bounded decision variables are often parameterized using sigmoid mappings to enforce box constraints. However, the sigmoid gradient diminishes near the bounds, which may weaken the optimization signal when constraints become active. This is particularly undesirable in OPF, where economically optimal solutions often lie on the boundary of the feasible region. We therefore adopt the following constraint-aware projection:
\begin{equation}
\mathcal P(\widetilde{\mathbf u})=\operatorname{Proj}_{[0,1]}(\widetilde{\mathbf u})=\min\{1,\max\{0,\widetilde{\mathbf u}\}\}.
\label{eq:normalized_projection}
\end{equation}

Accordingly, the voltage components are mapped to their admissible ranges by
\begin{equation} V_{i,t}=\underline{V}_{i}+\left(\overline{V}_{i}-\underline{V}_{i}\right)\mathcal P\left(\widetilde{V}_{i,t}\right). \label{eq:voltage_parameterization} \end{equation}

Similarly, the active-power components are parameterized recursively to enforce both generation and ramping limits. Given $P_{i,t-1}^{g}$, the admissible interval at time $t$ is
\begin{subequations}
\label{eq:recursive_bounds}
\begin{align}
\underline{P}_{i,t}^{\mathrm{adm}}&=\max\left\{\underline{P}_{i}^{g},P_{i,t-1}^{g}-R_{i}^{\mathrm{down}}\right\}, \label{eq:recursive_lower_bound}\\
\overline{P}_{i,t}^{\mathrm{adm}}&=\min\left\{\overline{P}_{i}^{g},P_{i,t-1}^{g}+R_{i}^{\mathrm{up}}\right\}. \label{eq:recursive_upper_bound}
\end{align}
\end{subequations}
The corresponding active-power dispatch is then obtained as
\begin{equation} P_{i,t}^{g}=\underline{P}_{i,t}^{\mathrm{adm}}+\left(\overline{P}_{i,t}^{\mathrm{adm}}-\underline{P}_{i,t}^{\mathrm{adm}}\right)\mathcal P\left(\widetilde{P}_{i,t}^{g}\right). \label{eq:pg_parameterization} \end{equation}

Collecting the above transformations gives
\begin{equation}
\label{eq:generator_composition}
\mathbf u=\mathcal F_{\mathrm{adm}}\left(\widetilde{\mathbf u}\right), ~
G_{\sigma}^{u}=\mathcal F_{\mathrm{adm}}\circ\widetilde{G}_{\sigma}^{u}
\end{equation}
where $\mathcal F_{\mathrm{adm}}(\cdot)$ denotes the deterministic constraint-preserving projection mapping. Consequently, the CSNG only needs to account for the remaining constraints associated with the dependent variables recovered through $\boldsymbol{\psi}_{\mathrm{pf}}(\cdot)$.

Although the projection in \eqref{eq:normalized_projection} allows the
generator to attain the variable bounds exactly, conventional clipping
blocks gradient propagation once the raw output enters a saturated region.
Specifically, let
\begin{equation} \mathbf u=\operatorname{Proj}_{[0,1]}(\widetilde{\mathbf u}),\quad g_\mathbf u=\frac{\partial\mathcal L}{\partial \mathbf u} \label{eq:projected_variable_gradient} \end{equation}
where $g_q$ denotes the upstream gradient of the current training objective $\mathcal L$.  Under the standard clipping operator, the gradient with respect to the raw output is
\begin{equation}
\label{eq:standard_clip_backward}
\left.
\frac{\partial \mathcal L}{\partial \widetilde{\mathbf u}}
\right|_{\mathrm{clip}}
=\frac{\partial \mathcal L}{\partial \mathbf u}
\frac{\partial \mathbf u}{\partial \widetilde{\mathbf u}}=
\begin{cases}
g_\mathbf u, & 0<\widetilde{\mathbf u}<1\\[1mm]
0, & \widetilde{\mathbf u}<0 \ \text{or}\ \widetilde{\mathbf u}>1
\end{cases}
\end{equation}

Hence, once $\widetilde{\mathbf u}$ enters the saturated region, informative gradient signals are blocked during backpropagation. To address this issue, we customize its backward propagation as follows:
\begin{equation} \left.
\frac{\partial \mathcal L}{\partial \widetilde{\mathbf u}}
\right|_{\mathrm{clip}} =\begin{cases} g_\mathbf u, & 0<\mathbf u<1,\\[1mm] g_\mathbf u, & \mathbf u=0\ \text{and}\ g_\mathbf u<0,\\[1mm] g_\mathbf u, & \mathbf u=1\ \text{and}\ g_\mathbf u>0,\\[1mm] 0, & \text{otherwise}. \end{cases} \label{eq:one_sided_projected_backward} \end{equation}

The rule has a simple interpretation. At the lower bound $\mathbf u=0$, $g_u<0$ is retained because the corresponding gradient-descent step increases $\mathbf u$, whereas $g_u>0$ is blocked because it would move the variable further outside the feasible interval. The
upper-bound case is symmetric. Thus, the customized rule preserves the exact
forward projection while allowing a saturated variable to leave the boundary
whenever the training objective favors an interior solution.

\subsection{Feasibility-Aware Self-Supervised End-to-End Training}
\label{subsec:optimization_training}

The CSNG is trained to generate a diverse set of candidate dispatch trajectories concentrated in feasible regions with low operating costs. A conventional supervised-learning approach would require a large number of labeled samples obtained by repeatedly solving the original SMPC problem in \eqref{eq:smpc_acopf} for different uncertainty instances, which would incur substantial computational cost. We therefore train the CSNG in a self-supervised manner by directly evaluating the operational constraints and objective function of the SMPC model on the generated candidates. For each training instance, $K_{\mathrm{tr}}$ candidate dispatch trajectories are sampled as
\begin{equation}
\mathbf u^{(k)}=G_{\sigma}^{u}\left(\mathbf \xi_{\mathrm{pool}},\mathbf z^{(k)}\right),
~ \mathbf z^{(k)}\overset{\mathrm{i.i.d.}}{\sim}\mathcal N(\mathbf 0,\mathbf I),
~ k=1,\ldots,K_{\mathrm{tr}}. \label{eq:training_candidates} \end{equation}

Each free-variable candidate is completed to the full operating point $\mathbf y^{(k)}$ through \eqref{eq:y_completion}. The resulting inequality-constraint residuals and operating costs are then used to train CSNG.

\subsubsection{Constraint Violation}
\label{subsubsec:hierarchical_violation}

A direct average over all inequality constraints may dilute the gradients associated with a small number of critical violations, particularly in large-scale SMPC problems involving many constraints, time periods, and uncertainty scenarios. To address this issue, we adopt a hierarchical mean--CVaR aggregation that preserves dense gradient information while emphasizing the most critical violations.

Let $\mathcal C$ denote the set of inequality-constraint families that remain after applying the admissible-output mapping $\mathcal F_{\mathrm{adm}}$, including generator reactive-power limits, reference-bus power limits, voltage limits, phase-angle-difference limits, and transmission-line thermal limits. For candidate $k$, scenario $s$, time period $t$, and constraint family $c\in\mathcal C$, define the normalized positive residual vector of \eqref{eq:scenario_value_ineq} as
\begin{equation} \mathbf r_{k,s,t}^{(c)}=\left[\mathbf D_c\mathbf g_{s,t}^{(c)}\left(\mathbf y^{(k)}\right)\right]_+\label{eq:group_normalized_residual} \end{equation}
where $\mathbf D_c$ is a diagonal scaling matrix that normalizes the numerical scales of different physical constraints. For a nonnegative vector $\mathbf r$, let $\operatorname{CVaR}_{\rho_c}(\mathbf r)$ denote the empirical mean of its largest $\rho_c$ fraction of elements. The violation associated with constraint family $c$ is defined as
\begin{equation} \mathcal V_{k,s,t}^{(c)}=\alpha_c\frac{\left\|\mathbf r_{k,s,t}^{(c)}\right\|_1}{m_c}+(1-\alpha_c)\operatorname{CVaR}_{\rho_c}\left(\mathbf r_{k,s,t}^{(c)}\right) \label{eq:local_constraint_violation} \end{equation}
where $m_c$ is the number of scalar inequalities in family $c$, and
$\alpha_c\in(0,1)$ controls the tradeoff between dense mean-violation
feedback and emphasis on the upper-tail violations.

The resulting feasibility loss over the generated candidate dispatch trajectories is
\begin{equation} 
\mathcal L_{\mathrm{fea}}=\sum_{k=1}^{K_{\mathrm{tr}}}
\sum_{s\in\mathcal N_s}
\sum_{t\in\mathcal T}
\sum_{c\in\mathcal C} \mathcal V_{k,s,t}^{(c)}. 
\label{eq:feasibility_shaping_loss}
\end{equation}

\subsubsection{Diversity Preservation}
\label{subsubsec:diversity_preservation}
Generating multiple candidates is useful only if different latent
samples can lead to distinct operating points. Excessive concentration
of the generated candidates in a narrow region would reduce the benefit
of stochastic sampling. We therefore introduce a diversity term among
candidates that have small constraint violations. The free-variable vector contains quantities with different physical ranges. To obtain a meaningful distance measure, each component is normalized according to its operating range:
\begin{equation}
\widehat{u}^{(k)}
=\frac{u^{(k)}-u^{\mathrm c}}{\Delta u},
\quad
u^{\mathrm c}=\frac{\overline{u}+\underline{u}}{2},
\quad
\Delta u=\frac{\overline{u}-\underline{u}}{2}
\label{eq:u_normalization}
\end{equation}
where $\underline u$ and $\overline u$ denote the corresponding lower and upper admissible limits.

For feasibility-aware distribution shaping, each candidate is assigned
a soft feasibility score 
\begin{equation} s_k=\exp\left(-\frac{\sum_{s\in\mathcal N_s}
\sum_{t\in\mathcal T}
\sum_{c\in\mathcal C} \mathcal V_{k,s,t}^{(c)}}{\tau_f}\right),~ 0<s_k\leq1\label{eq:soft_feasibility_score} \end{equation}
where $\tau_f>0$ determines the sensitivity to constraint violations. Candidates satisfying or closely approaching the operational constraints have $s_k$ close to one, whereas candidates with large violations have small scores. The score is used as a fixed feasibility-dependent weight in the distribution-shaping terms below.

To prevent different latent samples from collapsing onto nearby operating points, we extend the kernel-based repulsive regularization of \cite{wang2019improving} by incorporating the proposed soft feasibility score. Let $D$ denote the total trajectory dimension. The feasibility-weighted diversity loss is defined as
\begin{equation}
\mathcal{L}_{\mathrm{div}}=\frac{\displaystyle\sum_{i<j}\operatorname{sg}(s_i s_j)\exp\left(-\frac{\left\|\widehat{\mathbf u}^{(i)}-\widehat{\mathbf u}^{(j)}\right\|_2^2}{2D\sigma_d^2}\right)}{\displaystyle\sum_{i<j}\operatorname{sg}(s_i s_j)+\varepsilon}
\label{eq:diversity_loss}
\end{equation}
where $\sigma_d>0$ specifies the characteristic separation between candidates, $\varepsilon>0$ is a small constant for numerical stability, and $\operatorname{sg}(\cdot)$ denotes the stop-gradient operator. The Gaussian kernel is large for two nearby operating points and decreases as their distance increases. Therefore, minimizing \eqref{eq:diversity_loss} discourages near-duplicate candidates.

The factors $s_i s_j$ make this separation mainly act on candidate
pairs that are close to satisfying the operating constraints. For
example, if $s_1=0.95$, $s_2=0.90$, and $s_3=0.02$, the pairwise
weights are $s_1s_2=0.855$ and $s_1s_3=0.019$, respectively.
Consequently, two near-feasible candidates are encouraged to represent
different operating points, whereas a strongly infeasible candidate
receives little diversity-related force. The stop-gradient operation
makes $s_i s_j$ act only as a fixed weight. Otherwise, the diversity
loss could also be reduced by decreasing $s_k$, which corresponds to
increasing the constraint violation of a candidate.

\subsubsection{Economic Shaping}
\label{subsubsec:economic_shaping}
The economic objective should promote low-cost and feasible candidates. Directly minimizing generation cost may favor infeasible solutions with artificially low costs. Since inference selects the lowest-cost feasible candidate, greater emphasis is placed on the most competitive candidates. Accordingly, the soft feasibility score $s_k$ in \eqref{eq:soft_feasibility_score} is incorporated with a mean--best-$K_b$ aggregation:
\begin{equation}
\begin{aligned}
&\mathcal L_{\mathrm{eco}}={}\alpha_{\mathrm{eco}}\frac{1}{K_{\mathrm{tr}}}\sum_{k=1}^{K_{\mathrm{tr}}}C_{\mathrm{obj}}\left(\mathbf y^{(k)}\right)\left[1+\gamma_{\mathrm{eco}}\left(1-s_k\right)\right]\\
&+\left(1-\alpha_{\mathrm{eco}}\right)\frac{1}{K_b}\sum_{k\in\mathcal I_b}C_{\mathrm{obj}}\left(\mathbf y^{(k)}\right)\left[1+\gamma_{\mathrm{eco}}\left(1-s_k\right)\right]
\end{aligned}
\label{eq:economic_objective}
\end{equation}
where $\gamma_{\mathrm{eco}}>0$ controls the feasibility-dependent amplification and $\alpha_{\mathrm{eco}}\in(0,1)$ balances the contributions of the full candidate set and the best-$K_b$ subset. The index set $\mathcal I_b$ contains the $K_b$ candidates with the smallest feasibility-adjusted objective values $C_{\mathrm{obj}}(\mathbf y^{(k)})[1+\gamma_{\mathrm{eco}}(1-s_k)]$. As $s_k$ decreases with increasing constraint violation, infeasible candidates receive a larger economic penalty. The first term improves the overall economic quality of the generated distribution, while the second focuses optimization on the most competitive candidates likely to be selected at inference.

The feasibility loss reduces constraint violations, while the diversity loss discourages near-duplicate solutions. The economic shaping further promotes low-cost candidates. Together, these terms guide the stochastic generator toward diverse, feasible, and economically competitive dispatch trajectories. The complete CSNG training objective is therefore given by
\begin{equation} \mathcal L_{\mathrm{train}}=\lambda_{\mathrm{fea}}\mathcal L_{\mathrm{fea}}+\lambda_{\mathrm{div}}\mathcal L_{\mathrm{div}}+\lambda_{\mathrm{eco}}\mathcal L_{\mathrm{eco}} \label{eq:overall_training_loss} \end{equation}
where $\lambda_{\mathrm{fea}}$, $\lambda_{\mathrm{div}}$, and $\lambda_{\mathrm{eco}}$ balance the numerical scales and relative contributions of the three training terms.

To improve training stability, a two-stage training strategy is adopted. During the first stage, the CSNG is trained using $\mathcal L_{\mathrm{fea}}$ and $\mathcal L_{\mathrm{div}}$ to establish a feasible and non-collapsed candidate distribution. During the second stage, $\mathcal L_{\mathrm{eco}}$ is activated while the feasibility and diversity terms remain active, thereby shifting the generated distribution toward lower-cost regions without sacrificing feasibility or diversity. The complete training procedure is summarized in Algorithm~\ref{alg:self_supervised_node_training}.

\begin{algorithm}[htbp]
\caption{Two-Stage Self-Supervised End-to-End Training of the Conditional Stochastic Generator}
\label{alg:self_supervised_node_training}
\KwIn{training set $\mathcal D$; CSNG parameters $\sigma$; number of generated candidates $K_{\mathrm{tr}}$; stage boundary $E_1$; total epochs $E$; learning rate $\eta$}
\KwOut{trained CSNG parameters $\sigma$}
\SetAlgoLined
Initialize the CSNG parameters $\sigma$\;
\For{$e=1:E$}{
Sample a mini-batch of SMPC uncertainty instances $\boldsymbol{\xi}$ from $\mathcal D$\;
\ForEach{$\boldsymbol{\xi}$ in the mini-batch}{
Construct the pooled representation $\boldsymbol{\xi}_{\mathrm{pool}}$\;
Sample $\mathbf z^{(k)}\overset{\mathrm{i.i.d.}}{\sim}\mathcal N(\mathbf 0,\mathbf I)$ for $k=1,\ldots,K_{\mathrm{tr}}$\;
Generate candidate dispatch trajectories $\mathbf u^{(k)}=G_{\sigma}^{u}(\boldsymbol{\xi}_{\mathrm{pool}},\mathbf z^{(k)})$\;
Recover the corresponding operating points $\mathbf y^{(k)}$ through \eqref{eq:y_completion}\;

Compute $\mathcal L_{\mathrm{fea}}$ using \eqref{eq:group_normalized_residual}--\eqref{eq:feasibility_shaping_loss}\;
Compute $\mathcal L_{\mathrm{div}}$ using \eqref{eq:u_normalization}--\eqref{eq:diversity_loss}\;
\eIf{$e\leq E_1$}{
Set $\mathcal L=\lambda_{\mathrm{fea}}\mathcal L_{\mathrm{fea}}+\lambda_{\mathrm{div}}\mathcal L_{\mathrm{div}}$\;
}{
Compute $\mathcal L_{\mathrm{eco}}$ using \eqref{eq:economic_objective}\;
Set $\mathcal L=\lambda_{\mathrm{fea}}\mathcal L_{\mathrm{fea}}+\lambda_{\mathrm{div}}\mathcal L_{\mathrm{div}}+\lambda_{\mathrm{eco}}\mathcal L_{\mathrm{eco}}$\;
}
}
Update $\sigma\leftarrow\sigma-\eta\nabla_{\sigma}\mathcal L$\;
}
\end{algorithm}

\subsection{Differentiable Equality Completion Surrogate}
\label{subsec:pf_surrogate}
All operations in Algorithm~\ref{alg:self_supervised_node_training} are compatible with standard automatic differentiation except for the equality-completion mapping in \eqref{eq:y_completion}, which requires solving a nonlinear AC PF problem to recover the dependent variables. Although its sensitivities can be obtained through implicit differentiation \cite{donti2021dc3}, this is costly for large-scale scenario-based training. Each generated candidate requires converged PF solutions across all scenarios and time periods, followed by Jacobian-based linear solves during backpropagation. As a result, both computational and memory costs grow rapidly with the numbers of scenarios and candidates. To remove the nonlinear PF solver from the end-to-end CSNG training loop, we introduce a differentiable equality completion surrogate (DECS). The key idea is to learn only the implicit mapping from the standard PF specifications to the minimum set of unknown voltage states. Once these states are predicted, all remaining dependent quantities are recovered analytically through the AC network equations. Consequently, the learned component is restricted to the genuinely implicit part of the PF mapping, whereas the remaining physical relationships are retained exactly as deterministic differentiable operations.

\subsubsection{PF Specification and Minimal-State Mapping}
\label{subsubsec:decs_mapping}
Let $\mathcal N_{\mathrm{PV}}$ and $\mathcal N_{\mathrm{PQ}}$ denote the sets of P-V and P-Q buses, respectively, where the reference bus $r_0$ is excluded from $\mathcal N_{\mathrm{PV}}$. Let $\boldsymbol{\rho}^{\mathrm{PV}}$ and $\boldsymbol{\rho}^{\mathrm{PQ}}$ denote the PF specifications associated with the P-V and P-Q buses, respectively, and let $\boldsymbol{\rho}$ denote the complete PF specification vector. Specifically,
\begin{subequations}
\label{eq:decs_pf_input}
\begin{align}
\boldsymbol{\rho}^{\mathrm{PV}}
&=\left(\mathbf p_{\mathrm{PV}}^{\mathrm{sp}},\mathbf V_{\mathrm{PV}}\right)
\label{eq:decs_pf_input_pv}
\\
\boldsymbol{\rho}^{\mathrm{PQ}}
&=\left(\mathbf p_{\mathrm{PQ}}^{\mathrm{sp}},\mathbf q_{\mathrm{PQ}}^{\mathrm{sp}}\right)
\label{eq:decs_pf_input_pq}
\\
\boldsymbol{\rho}
&=\left(\boldsymbol{\rho}^{\mathrm{PV}},V_{r_0},\boldsymbol{\rho}^{\mathrm{PQ}}\right)
\label{eq:decs_pf_input_full}
\end{align}
\end{subequations}
where $\mathbf p_{\mathrm{PV}}^{\mathrm{sp}}$ denotes the vector of net active-power injections at the P-V buses, $\mathbf V_{\mathrm{PV}}$ denotes the corresponding voltage magnitudes, and $\mathbf p_{\mathrm{PQ}}^{\mathrm{sp}}$ and $\mathbf q_{\mathrm{PQ}}^{\mathrm{sp}}$ denote the vectors of net active- and reactive-power injections at the P-Q buses, respectively. The quantity $V_{r_0}$ denotes the voltage magnitude at the reference bus, whose phase angle is fixed as $\theta_{r_0}=0$.

In the proposed SMPC framework, $\boldsymbol{\rho}$ is determined jointly by the free variables $\mathbf u$ and the uncertainty realization $\boldsymbol{\xi}$, which is compactly expressed as
$\boldsymbol{\rho}=\boldsymbol{\rho}\left(\mathbf u;\boldsymbol{\xi}\right)$. Given $\boldsymbol{\rho}$, let $\boldsymbol{\chi}$ denote the minimal set of
unknown PF states,
\begin{equation}
\boldsymbol{\chi}
=
\left(
\boldsymbol{\theta}_{\mathrm{PV}},
\boldsymbol{\theta}_{\mathrm{PQ}},
\mathbf V_{\mathrm{PQ}}
\right).
\label{eq:decs_pf_state}
\end{equation}

The AC PF equations implicitly define the mapping $\boldsymbol{\chi}
=\boldsymbol{\phi}_{\mathrm{pf}}\left(\boldsymbol{\rho}\right)$. The proposed DECS replaces the repeated numerical evaluation of $\boldsymbol{\phi}_{\mathrm{pf}}(\cdot)$ with a differentiable surrogate $\Phi_{\omega}(\cdot)$:
\begin{equation}
\widehat{\boldsymbol{\chi}}
=
\Phi_{\omega}
\left(
\boldsymbol{\rho}
\right),
\label{eq:decs_neural_mapping}
\end{equation}
where $\omega$ denotes the trainable parameters of the surrogate.

\subsubsection{Physics-Guided Training}
\label{subsubsec:decs_training}
The DECS is pretrained offline and is kept fixed during the subsequent self-supervised training of the CSNG. The surrogate training set is constructed by sampling PF specifications over the operating region relevant to the SMPC problem. For each sampled PF specification $\boldsymbol{\rho}^{(n)}$, an AC PF solver is executed once offline to obtain the corresponding solution $\boldsymbol{\chi}^{(n)}$. The resulting data set is
\begin{equation} \mathcal D_{\mathrm{pf}}=\left\{\left(\boldsymbol{\rho}^{(n)},\boldsymbol{\chi}^{(n)}\right)\right\}_{n=1}^{N_{\mathrm{pf}}}. \label{eq:decs_dataset} \end{equation}

A purely data-driven supervised loss can be defined as 
\begin{equation} 
\mathcal L_{\mathrm{sup}}=\frac{1}{N_{\mathrm{pf}}}\sum_{n=1}^{N_{\mathrm{pf}}}\|\widehat{\boldsymbol{\chi}}^{(n)}-\boldsymbol{\chi}^{(n)}\|_2^2.
\label{eq:decs_supervised_loss} 
\end{equation}

However, a purely data-driven supervised loss does not explicitly capture the underlying physical relationships governed by the PF equations, which may limit the generalization capability of the model. To enhance generalization, we therefore augment the supervised objective with an explicit AC-physics loss. For compactness, define the AC nodal-injection functions in \eqref{eq:smpc_acopf_pbal} and \eqref{eq:smpc_acopf_qbal} as $\mathcal P_i(\mathbf V,\boldsymbol{\theta})$ and $\mathcal Q_i(\mathbf V,\boldsymbol{\theta})$. For a predicted state $\widehat{\boldsymbol{\chi}}^{(n)}$, the complete voltage vector is assembled using the specified P-V/reference-bus voltage magnitudes and the predicted P-Q voltage magnitudes. Denote the resulting voltage state by $(\widehat{\mathbf V}^{(n)},\widehat{\boldsymbol{\theta}}^{(n)})$. The active-power residuals at all nonreference buses and the reactive-power residuals at the P-Q buses are then
\begin{subequations}
\label{eq:decs_pf_residual}
\begin{align}
\mathbf r_P^{(n)}&=\mathcal P_{\mathcal N_{\mathrm{PV}}\cup\mathcal N_{\mathrm{PQ}}}\left(\widehat{\mathbf V}^{(n)},\widehat{\boldsymbol{\theta}}^{(n)}\right)-\mathbf p_{\mathcal N_{\mathrm{PV}}\cup\mathcal N_{\mathrm{PQ}}}^{\mathrm{sp},(n)}, \label{eq:decs_active_residual}\\
\mathbf r_Q^{(n)}&=\mathcal Q_{\mathcal N_{\mathrm{PQ}}}\left(\widehat{\mathbf V}^{(n)},\widehat{\boldsymbol{\theta}}^{(n)}\right)-\mathbf q_{\mathcal N_{\mathrm{PQ}}}^{\mathrm{sp},(n)}. \label{eq:decs_reactive_residual}
\end{align}
\end{subequations}

No reactive-power residual is imposed at a P-V bus, and no active- or reactive-power residual is imposed at the reference bus, because these quantities are dependent PF outputs rather than specified inputs. The physics-guided loss and the resulting DECS training objective are defined as
\begin{subequations}
\label{eq:decs_training_objective}
\begin{align}
\mathcal L_{\mathrm{phy}}&=\frac{1}{N_{\mathrm{pf}}}\sum_{n=1}^{N_{\mathrm{pf}}}\left[\left\|\mathbf r_P^{(n)}\right\|_2^2+\left\|\mathbf r_Q^{(n)}\right\|_2^2\right]\label{eq:decs_physics_loss}\\
\omega^{\star}&=\arg\min_{\omega}\left(\mathcal L_{\mathrm{sup}}+\lambda_{\mathrm{phy}}\mathcal L_{\mathrm{phy}}\right)\label{eq:decs_training_optimum}
\end{align}
\end{subequations}
where $\lambda_{\mathrm{phy}}>0$ controls the relative emphasis placed on physical consistency.

\subsubsection{Exact Differentiable AC Reconstruction}
\label{subsubsec:decs_reconstruction}
The NN in \eqref{eq:decs_neural_mapping} predicts only the minimal voltage-state vector. All remaining PF-dependent quantities are subsequently recovered from the AC equations through a parameter-free reconstruction layer. For each scenario $s$ and time period $t$, the full voltage magnitudes and phase angles are first assembled as
\begin{subequations}
\label{eq:decs_voltage_assembly}
\begin{align}
\widehat V_{s,i,t}&=\begin{cases}V_{s,i,t},&i\in\mathcal N_{\mathrm{PV}}\cup\{r_0\},\\[1mm]\left[\widehat{\boldsymbol{\chi}}_{s,t}\right]_{V_i},&i\in\mathcal N_{\mathrm{PQ}},\end{cases} \label{eq:decs_voltage_magnitude_assembly}\\
\widehat\theta_{s,i,t}&=\begin{cases}0,&i=r_0,\\[1mm]\left[\widehat{\boldsymbol{\chi}}_{s,t}\right]_{\theta_i},&i\in\mathcal N_{\mathrm{PV}}\cup\mathcal N_{\mathrm{PQ}}.\end{cases} \label{eq:decs_voltage_angle_assembly}
\end{align}
\end{subequations}
Using these voltages, the active- and reactive-power outputs of each conventional generator is reconstructed as
\begin{subequations}
\label{eq:decs_slack_reconstruction}
\begin{align}
\widehat P^g_{s,i,t}&=\mathcal P_{_i}\left(\widehat{\mathbf V}_{s,t},\widehat{\boldsymbol{\theta}}_{s,t}\right)-P^r_{s,_i,t}+P^d_{s,_i,t}, \label{eq:decs_slack_p_reconstruction}\\
\widehat Q^g_{s,i,t}&=\mathcal Q_i\left(\widehat{\mathbf V}_{s,t},\widehat{\boldsymbol{\theta}}_{s,t}\right)-Q^r_{s,i,t}+Q^d_{s,i,t}.\label{eq:decs_slack_q_reconstruction}
\end{align}
\end{subequations}

The branch flows $\widehat P_{s,ij,t}$ and $\widehat Q_{s,ij,t}$, together with the apparent-power flow $\widehat S_{s,ij,t}$, are then evaluated directly from \eqref{eq:smpc_acopf_pflow}--\eqref{eq:smpc_acopf_sflow} using the reconstructed voltage states $(\widehat{\mathbf V}_{s,t},\widehat{\boldsymbol{\theta}}_{s,t})$.

Collecting the predicted voltage states and all analytically reconstructed quantities gives
\begin{equation} \widehat{\mathbf w}_{s,t}=\mathcal R_{\mathrm{AC}}\left(\boldsymbol{\rho}_{s,t},\Phi_{\omega^\star}(\boldsymbol{\rho}_{s,t});\boldsymbol{\xi}_{s,t}\right)\label{eq:decs_reconstruction_operator} \end{equation}
where $\mathcal R_{\mathrm{AC}}(\cdot)$ denotes the AC reconstruction operator consisting only of the algebraic operations in \eqref{eq:decs_voltage_assembly} and \eqref{eq:decs_slack_reconstruction}. The resulting DECS mapping can therefore be expressed as
\begin{subequations}
\label{eq:decs_complete_mapping}
\begin{align}
\boldsymbol{\rho}_{s,t}&=\boldsymbol{\rho}\left(\mathbf u_t;\boldsymbol{\xi}_{s,t}\right), \label{eq:decs_complete_mapping_input}\\
\widehat{\boldsymbol{\chi}}_{s,t}&=\Phi_{\omega^\star}\left(\boldsymbol{\rho}_{s,t}\right), \label{eq:decs_complete_mapping_state}\\
\widehat{\boldsymbol{\psi}}_{\mathrm{pf}}\left(\mathbf u_t;\boldsymbol{\xi}_{s,t}\right)&=\mathcal R_{\mathrm{AC}}\left(\boldsymbol{\rho}_{s,t},\widehat{\boldsymbol{\chi}}_{s,t};\boldsymbol{\xi}_{s,t}\right). \label{eq:decs_complete_mapping_output}
\end{align}
\end{subequations}

Thus, during CSNG training, the original non-differentiable equality-completion operation in \eqref{eq:y_completion} is replaced by the following differentiable operator:
\begin{equation} \widehat{\mathbf y}^{(k)}=\begin{bmatrix}\mathbf u^{(k)}\\[1mm]\widehat{\boldsymbol{\psi}}_{\mathrm{pf}}\left(\mathbf u^{(k)};\boldsymbol{\xi}\right)\end{bmatrix}. \label{eq:decs_y_completion} \end{equation}

After training, the CSNG solves \eqref{eq:compact_formulation} online through a generate--evaluate--select procedure. For a new SMPC instance $\boldsymbol{\xi}$, the pooled representation $\boldsymbol{\xi}^{\mathrm{pool}}$ and $K$ latent samples $\mathbf z^{(k)}$ are used to generate $K$ candidate trajectories
$\mathbf u^{(k)}=G_{\sigma}^{u}(\boldsymbol{\xi}^{\mathrm{pool}},\mathbf z^{(k)})$
. The remaining states are recovered using the original AC PF equations, and each candidate is checked against the SMPC constraints. The lowest-cost feasible candidate is selected as the final dispatch solution. Thus, online SMPC solution reduces to parallel candidate generation, PF-based feasibility verification, and best-feasible selection.

\section{Case Study}
\label{sec:case}
To evaluate the effectiveness of the proposed method, numerical experiments are conducted on the IEEE 14-bus and IEEE 118-bus systems \cite{pglib_opf}. The intraday scheduling horizon consists of 16 time periods, each with a duration of 15 minutes, with 20 uncertainty scenarios considered for each instance. A total of 5,000 uncertainty instances are generated for CSNG training, with the uncertain quantities sampled within $\pm 15\%$ of their nominal values. The datasets are divided into training, validation, and test sets with a ratio of $8{:}1{:}1$. The training set is used for parameter optimization, the validation set for early stopping and overfitting control, and the test set for final performance evaluation. PF calculations are performed using \texttt{Power Grid Model} \cite{power-grid-model}, which supports parallel PF computations, while the NNs are implemented in \texttt{PyTorch}. All CPU-based computations are performed on an Intel i9-14900HX processor, while GPU-based neural-network training is conducted on an NVIDIA RTX 5090 GPU.

The proposed method is compared with the following approaches: 
(i) direct solution of the nonconvex SMPC problem using IPOPT, denoted as \textbf{IPOPT}; 
(ii) a deterministic NN trained with a penalty-based loss \cite{11192606}, denoted as \textbf{D-NN};
(iii) the proposed stochastic generative architecture with sigmoid-based output parameterization, denoted as \textbf{S-CSNG}; 
(iv) the proposed stochastic generative architecture without the diversity-preservation term, denoted as \textbf{WD-CSNG}; and 
(v) the complete proposed stochastic generative method, denoted as \textbf{CSNG}.
At inference, all candidates generated by CSNG are evaluated in parallel using \texttt{Power Grid Model} \cite{power-grid-model}, and the feasible candidate with the lowest operating cost is selected as the final solution. All data, hyperparameter settings, and implementation source codes are available at \url{https://github.com/JieZhu6/Generative_SMPC}.

\subsection{Case Study on the IEEE 14-Bus System}
For the IEEE 14-bus system, 10,000 PF operating points are generated to train DECS. For each SMPC instance, the CSNG generates 50 candidate dispatch trajectories.

\subsubsection{Accuracy of DECS}
\begin{figure}[!t]
   \centering
   \includegraphics[width=0.47\textwidth]{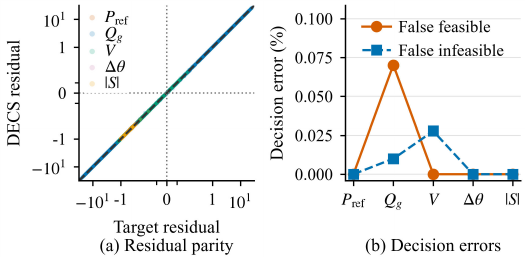}
   \caption{Approximation Accuracy of DECS on the IEEE 14-Bus System}
   \label{fig:decs_validation_14}
\end{figure}

The accuracy of DECS is evaluated on the held-out validation set, as shown in Fig.~\ref{fig:decs_validation_14}, using the normalized signed residuals of five inequality-constraint groups: reference-generator active-power limits, generator reactive-power limits, P-Q bus voltage limits, branch angle-difference limits, and thermal limits at both branch ends. The mean absolute residual errors range from $7.77\times10^{-5}$ to $2.97\times10^{-3}$, while the largest 95th-percentile error remains below $9.40\times10^{-3}$. The feasibility-classification agreement rates range from $99.92\%$ to $100\%$, with maximum false-feasible and false-infeasible rates of only $0.07\%$ and $0.028\%$, respectively. These results indicate that DECS provides sufficiently accurate constraint evaluations for offline CSNG training.

It should be noted that DECS is used only during offline training to provide differentiable constraint feedback. At inference, the generated dispatch candidates are validated using the original AC power-flow equations. Therefore, the small approximation errors of DECS do not affect the reliability of online decisions.

\subsubsection{Performance Comparison and Ablation Study}
\begin{figure}[!t]
   \centering
   \includegraphics[width=0.47\textwidth]{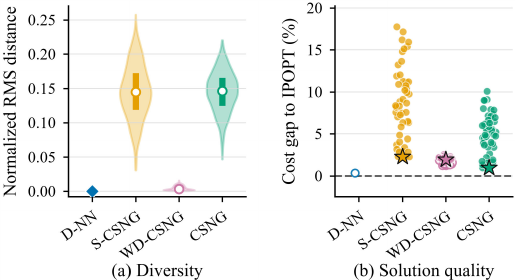}
   \caption{Dispatch Results on the IEEE 14-Bus System}
   \label{fig:dispatch_quality_14}
\end{figure}

As shown in Table~\ref{tab:com_14} and Fig.~\ref{fig:dispatch_quality_14}, the proposed CSNG achieves a favorable balance among feasibility, solution quality, and computational efficiency. Compared with the deterministic D-NN, whose feasibility rate is only $46.2\%$, CSNG achieves a $100\%$ feasibility rate by generating multiple candidate trajectories and selecting the lowest-cost feasible solution. Meanwhile, its objective value is 35,310, corresponding to an optimality gap of only $0.80\%$ relative to IPOPT. This result highlights the advantage of stochastic generation over single-shot prediction by providing multiple opportunities to recover a feasible and economical solution for each uncertainty instance.

Compared with S-CSNG, which uses sigmoid-based output parameterization, CSNG reduces the objective value from 35,658 to 35,310 and the optimality gap from $1.79\%$ to $0.80\%$. This improvement is attributed to the proposed constraint-aware projection, which allows decision variables to reach active bounds while preserving informative gradients through the customized backward rule, thereby mitigating the vanishing-gradient behavior of sigmoid mappings near active constraints.

The benefit of diversity preservation is illustrated in Fig.~\ref{fig:dispatch_quality_14}(a). Without the diversity term, WD-CSNG produces highly concentrated candidates, whereas CSNG maintains a larger normalized root mean square (RMS) distance and thus preserves distinct dispatch alternatives. As shown in Fig.~\ref{fig:dispatch_quality_14}(b), this diversity enables CSNG to retain a broad candidate set while still identifying low-cost feasible solutions. Consequently, CSNG achieves a $100\%$ feasibility rate and a lower objective value than WD-CSNG, with an average solution time of only 0.19~s per instance, approximately 100 times faster than IPOPT. Compared with D-NN, the additional online computation of CSNG arises from the PF-based validation of the generated candidates. Since these PF calculations are performed in parallel, the resulting overhead is acceptable for online operation.

\begin{table}
\caption{Method Comparison on the IEEE 14-Bus System}
\label{tab:com_14}
\centering
{
\begin{tabular}{ccccc}
\hline
Method& Obj. Value&Opt. Gap& Feas. Rate&Comp. Time\\
\hline
 IPOPT&  35031 &-& 100\%&19.86s\\
 D-NN&  35140 &0.31\%& 46.2\%&0.01s\\
S-CSNG& 35658 &1.79\%& 100\%& 0.19s\\
 WD-CSNG& 35362 &0.94\%& 99.8\%&0.19s\\
CSNG& 35310 &0.80\%&  100\%& 0.19s\\
\hline
\end{tabular}
}
\end{table}

\subsection{Case Study on the IEEE 118-bus System}
For the IEEE 118-bus system, 50,000 PF operating points are generated to train DECS. For each SMPC instance, the CSNG produces 100 candidate dispatch trajectories.

\subsubsection{Accuracy of DECS}
\begin{figure}[!t]
   \centering
   \includegraphics[width=0.47\textwidth]{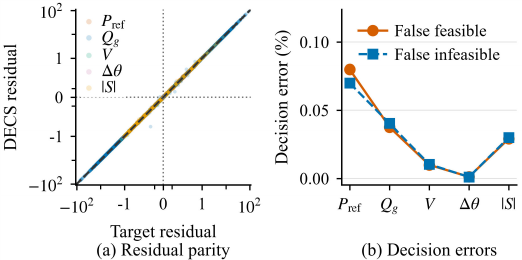}
   \caption{Approximation Accuracy of DECS on the IEEE 118-Bus System}
   \label{fig:decs_validation_118}
\end{figure}

The accuracy of the retrained DECS is evaluated on the held-out validation set, as shown in Fig.~\ref{fig:decs_validation_118}. The predicted residuals closely follow the identity line, with small approximation errors and feasibility-classification agreement rates above $99.85\%$. These results confirm that DECS provides sufficiently accurate differentiable constraint feedback for offline CSNG training.

\subsubsection{Performance Comparison and Ablation Study}
\begin{figure}[!t]
   \centering
   \includegraphics[width=0.47\textwidth]{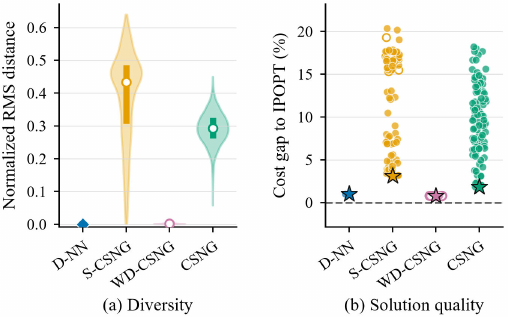}
   \caption{Dispatch Results on the IEEE 118-Bus System}
   \label{fig:dispatch_quality_118}
\end{figure}

As shown in Table~\ref{tab:com_118} and Fig.~\ref{fig:dispatch_quality_118}, CSNG achieves a $100\%$ feasibility rate, compared with $80\%$ for the deterministic D-NN, while increasing the optimality gap by only $0.81$ percentage points. This result demonstrates that multi-candidate generation substantially improves the reliability of feasible-solution recovery without sacrificing much economic performance. Compared with S-CSNG, CSNG reduces the optimality gap from $3.10\%$ to $1.81\%$, confirming the benefit of the proposed constraint-aware projection in preserving effective optimization signals near active bounds. Moreover, WD-CSNG achieves only a $36.2\%$ feasibility rate, indicating that the diversity-preservation loss is essential for preventing candidate collapse and maintaining sufficiently distinct dispatch alternatives.

In terms of computational efficiency, IPOPT requires nearly 400~s per instance on average for the IEEE 118-bus system, which is impractical for intraday dispatch. In contrast, CSNG retains a much lower online solution time while achieving full feasibility. Its current inference time is mainly determined by the available parallelism for PF-based candidate verification. Since these PF evaluations are mutually independent, sufficient parallel computing resources can further reduce the overall runtime, theoretically approaching the time required for a single PF calculation and potentially falling below 1~s.

\begin{table}
\caption{Method Comparison on the IEEE 118-Bus System}
\label{tab:com_118}
\centering
{
\begin{tabular}{ccccc}
\hline
Method & Obj. Value & Opt. Gap & Feas. Rate & Comp. Time\\
\hline
 IPOPT &  1,541,150 & - & 100\%& 398.02s\\
 D-NN &  1,556,493 & 1.00\%& 83\%&0.01s\\
S-CSNG & 1,588,919& 3.10\%& 100\% & 14.27s\\
 WD-CSNG & 1,552,132 & 0.71\%& 36.2\%& 16.17s\\
CSNG & 1,568,985 & 1.81\%  &  100\% & 13.74s\\
\hline
\end{tabular}
}
\end{table}

\subsubsection{Sensitivity Analysis}
\begin{figure}[!t]
   \centering
   \includegraphics[width=0.47\textwidth]{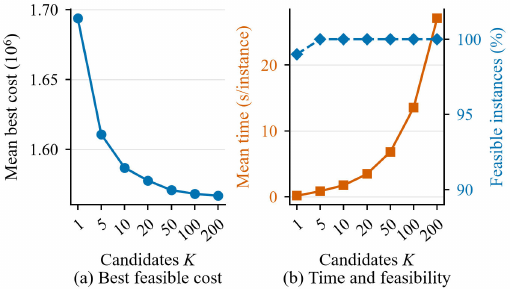}
   \caption{Sensitivity to the Number of Candidate Solutions}
   \label{fig:candidate_sensitivity_118}
\end{figure}

As shown in Fig.~\ref{fig:candidate_sensitivity_118}, increasing the number of candidates $K$ improves both the feasibility robustness and economic quality of the selected solution. As $K$ increases from 1 to 50, the average best cost over commonly feasible test instances decreases from $1.694\times10^6$ to $1.571\times10^6$, corresponding to a $7.26\%$ reduction, with the most pronounced improvement observed for $K=1$--10. Moreover, all test instances obtain at least one feasible candidate once $K\geq5$, highlighting the benefit of multi-candidate generation for reliable feasible-solution recovery. Further increasing $K$ yields diminishing economic returns; for example, increasing $K$ from 100 to 200 reduces the average best cost by only $0.088\%$ while increasing the computation time from 13.52~s to 27.08~s. Considering that the online computation time is primarily determined by the available parallel PF computing capacity, $K=100$ is adopted under the computing resources used in this study as a practical tradeoff among solution feasibility, operating economy, and computational efficiency.

\section{Conclusion}
\label{sec:con}
This paper develops an end-to-end generative framework for stochastic model predictive dispatch under nonlinear AC network constraints. By using stochastic generation for feasible-solution recovery, the proposed CSNG produces multiple dispatch candidates for each uncertainty realization and improves feasibility over deterministic single-output surrogates. A feasibility-aware self-supervised distribution-shaping strategy guides the candidates toward diverse, feasible, and economical regions without requiring optimal SMPC labels. Together with constraint-aware projection and differentiable equality completion, the framework enables efficient end-to-end training while preserving key operational constraints and AC network physics. Case studies on the IEEE 14- and 118-bus systems show that the proposed method consistently recovers feasible near-optimal solutions with substantial computational savings over conventional optimization.

\bibliographystyle{IEEEtran}
\bibliography{ref}

\end{document}